## RESEARCH

# Analyzing Chromatin Using Tiled Binned Scatterplot Matrices

Dirk Zeckzer[1,2], Daniel Gerighausen[1,3], Lydia Steiner[3,4*] and Sonja J. Prohaska[3]

### Abstract

**Background:** Over the last years, more and more biological data became available. Besides the pure amount of new data, also its dimensionality—the number of different attributes per data point—increased. Recently, especially the amount of data on chromatin and its modifications increased considerably. In the field of epigenetics, appropriate visualization tools designed for highlighting the different aspects of epigenetic data are currently not available.

**Results:** We present a tool called TiBi-Scatter enabling correlation analysis in 2D. This approach allows for analyzing multidimensional data while keeping the use of resources such as memory small. Thus, it is in particular applicable to large data sets.

**Conclusions:** TiBi-Scatter is a resource-friendly and easy to use tool that allows for the hypothesis-free analysis of large multidimensional biological data sets.

## Background

Over the last years, more and more biological data became available. Besides the pure amount of new data, also its dimensionality—the number of different attributes per data point—increased. Recently, especially the amount of data on chromatin and its modifications increased considerably. Although this is beneficial for the researchers of that field, who can base their findings on a solid basis, the same researchers face a considerable problem: analyzing this large amount of data becomes infeasible using currently available tools. This problem is aggravated by the lack of suitable visualization approaches. Besides the approach of [1] based on self-organizing maps (SOMs) only few visualizations targeted to chromatin are available.

Our goal is a tool for analyzing the patterns of epigenetic marks and their changes during differentiation. This necessitates, that the discovery of patterns in the transient epigenetic states should be enabled using our tool. Thus, a segmentation of the genome into modified and unmodified regions is required that is independent of annotations on the genome.

Our contributions are:
- *Visualization*
  - A scatterplot matrix visualization that was adapted to foster the analysis of chromatin: tiled-binned scatterplot matrices (TiBi-SPLOM).
- *Biology*
  - Analysis of the co-occurrence pattern of the studied modifications.
  - Data derived hypotheses on the recruitment mechanisms of H3K4me3 and the structure of the epigenome.

### Chromatin and Epigenetics

A human being consists of trillions of cells of several hundred cell types. Each cell, no matter which cell type, contains the same genome encoding all genes ever needed during life. However, depending on the cell type, only a part of the genes are expressed. Therefore, regulatory mechanism have to exist to ensure that in each cell only a well defined set of genes is expressed. A key role in regulation plays the epigenome. Furthermore, it plays a key role in differentiation—the switch from one progenitor cell type into a more specific cell type. Thus, more and more research is done in the field of epigenetics: "the study of heritable changes in gene expression that are not mediated at the DNA sequence level" [2].

### Basic Modules of (Epi)genetics

The DNA carries the genetic information and exists as two complementary strands of many nucleotides also called bases forming a double helix. It is further packed

*Correspondence: lydia@bioinf.uni-leipzig.de
[3]Computational EvoDevo, Department of Computer Science, Leipzig University, Härtelstraße 16-18, 04107 Leipzig, Germany
Full list of author information is available at the end of the article into nucleosomes consisting of a protein complex and 146 base pairs of DNA wrapped approximately 1.6 times around the protein complex. Two nucleosomes are separated by around 80 base pairs of DNA. The unity of protein complexes and DNA is called chromatin [2].

Chromatin exists in different structural modes. During cell division, the chromatin is strongly condensed into a compact transport form. It has the typical "X-shape" and is called metaphasen chromatin [3]. In this mode, no transcription occurs. Most of the time, chromatin exists in less condensed forms. Parts, which are only lightly packed are called euchromatin [2]. Due to the higher accessibility of the DNA, euchromatin is often highly transcribed. Tightly packed chromatin is called heterochromatin and results in gene silencing [2].

It was found that packaging depends on the state of the protein complex wrapping the DNA. The protein complex is an octamer consisting of histones—a special class of proteins. In a nucleosome, 2 copies of each of the 4 different types, i.e., H2A, H2B, H3, and H4, are incorporated. Each of them can be modified by adding chemical groups and small proteins, e.g., acetyl-groups or methyl-groups. While acetylation is mostly associated with high transcription rates, the effects of methylations are very diverse [4, 2, 3].

In this paper, three methylations are studied: H3K4me3, H3K27me3, and H3K9me3. Trimethylation (me3) at histone H3 at the lysine (one letter code K) at position 4 (H3K4me3) is associated with highly expressed genes [4]. It is abundantly found in embryonic stem cells and is very important in the early developmental stages. Genes that are expressed in all cells, so called housekeeping genes, and tissue-specific genes are often associated with histones carrying this modification.

A contrary effect has trimethylation at histone H3 at the lysine at position 27 (H3K27me3) [5]. It is found that it leads to heterochromatin formation. Histones can carry both marks, H3K4me3 and H3K27me3. This state is called poised chromatin and suppresses transcription. Nevertheless, in one step, i.e., removing the H3K27me3 mark, this state can be transformed into an active state. Therefore, it marks genes in undifferentiated cells that have to be transcribed in later differentiation stages [6].

The last modification considered in this paper is trimethylation at histone H3 at the lysine at position 9 (H3K9me3). Like H3K27me3, it is associated with heterochromatin and thus, represses gene expression [2]. It was found that it can recruit DNA methylases, which add methyl-groups to the DNA [7]. DNA methylation can hardly be reversed and efficiently suppresses transcription. Furthermore, it is inherited with high accuracy during cell division to the daughter cells.

Disregarding the different functions, all three modifications together as well as any combination of these marks can be observed in the genome [8]. Thus, analyzing the combinations of these marks and the changes during differentiation provides insights in the mechanisms leading to the modifications.

## Data Sets and Segmentation

ChIP-seq enables fast measurement of the genome-wide distribution of various epigenetic marks in multiple cell states. Experiments result in billions of sequences called reads representing the DNA associated with measured marks. This data has to be mapped back onto the genome, and genomic regions with sufficient read enrichment have to be identified.

For this contribution, the recently analyzed data set from [1] is used. It is based on the data set published in [6] dealing with mouse cells. It consists of the 3 modifications described before, i.e., H3K4me3, H3K27me3, and H3K9me3, measured in 3 cell types, i.e., embryonic stem cells (ESC), murine embryonic fibroblasts (MEF), and neuronal progenitor cells (NPC). The three cell types depict three different stages of differentiation. ESC are in the first stage and can differentiate into any cell type. MEF are already differentiated one stage further, but still can become a high number of different cell types. NPC are the highest differentiated cell type in this set and can only be differentiated into neuronal cell types. The data was measured using the ChIP-seq technique. Antibodies designed to bind a specific modification are used to select only those parts of the DNA that is wrapped around nucleosomes carrying the mark to measure. Since a lot of cell material is needed for ChIP-seq, the measurement is done using cell populations and not for single cells. Therefore, the results represent the population average rather than the modification in one cell. This means, the resulting signal is a mixture of the signals of all cells in the population. Another drawback of this method is that the accuracy requires an additional experiment measuring the background/noise of the method. Afterwards, only those parts of the genome are called modified, which are found to be sufficiently enriched compared to the background. This normalization is done for each of the experiments measuring a modification individually. Thus, even if the same region of the genome is modified with two different modifications, it is likely that the regions identified after normalization are slightly different.

The goal of the biologist is to compare the different modifications in the different cell types. This implies, that segmentations of the genome are needed that resolve the boundary problem as described above, i.e., that define, when two regions are the same and when they are different. We decided using a data-driven genome segmentation that uses one cell type as a reference and compares the modification data from the remaining cell types to the reference cell type. In order to do this, the boundaries of all modified regions in one cell type (regardless of the modification) are projected onto the genome. Two subsequent boundaries define the borders of a segment. Segments of a size smaller than 200bp (about the length of DNA associated with one nucleosome) were eliminated, since they likely arose from noise in the data, e.g., slightly different boundaries coming from different experiments. This segmentation is called ES-segmentation and is described in more detail in [1].

We calculated a 9-dimensional vector for each of the segments. Each dimension represents one modification in one cell type and indicates the percentage of the nucleotides of the region that is covered with this modification in the corresponding cell type. By construction, the coverage with the modification from the reference cell type has to be either 0 or 1, while the coverage by the modifications from the remaining cell types are values from the interval [0, 1]. Additional information can be calculated for each segment and is then attached to the vector. Thus, the vector describing the segments may increase in size but has the same dimensions for each of the data points.

For this contribution, we choose the ES-segmentation with the ESC as reference cell type. Thus, the boundaries of all regions from the three modifications in ESC define the borders of the segments. In the case of our data set, this results in more than 800.000 segments. Segment coverage by the three modifications in the three cell types were calculated as the fraction of the segments' length modified by the respective modification. Additionally, the CpG-density and the length of each segment is stored. Thus, each segment is described by an 11-dimensional vector representing the coverage in ESC, MEF, and NPC, the CpG-density, and the length of the segment. Due to the construction of the segments, the first 3 entries in the vector are either 0 or 1 (entries for ESC) and the entries for MEF and NPC (entries 4-9) are in the range of [0, 1]. Thus, the first 3 columns can be converted into a code (0-7) by interpreting them as a binary number. Entry 10 represents the CpG-density which in theory ranges between 0 and 1 but is found to be usually much smaller than 1 due to high mutation rate of the CG-dinucleotid. The last entry represents the length and is therefore the integer between 200 (minimal segment length) and, theoretically, the size of the largest chromosome in the data set.

## Previous Work

A large variety of methods was proposed to map multi-dimensional data onto 2D visual space. Icon-based approaches represent the attributes by using so called icons, like Chernoff faces [9] or stick figures [10]. In both approaches, attributes are not mapped uniformly to graphical elements, making some attribute values more prominent than others. Further, we can consider the tiles used in our approach as icons and thus get a clear distinction between attributes and categories in our approach.

Axis-based methods map each attribute onto one axis, select the attributes to display and then arrange the respective axes to allow the analysis of correlations or comparisons. These methods include scatterplot matrices, radar charts, parallel coordinate plots [11], Hyperboxes [12], and TimeWheels [13]. In general, there is not natural order for these axis for any of the proposed methods. Peng et al. [14] proposed a method for computing such an order based on the idea of clutter reduction. They propose a metric for the amount of clutter in a view and then order the axes such that the clutter is reduced. Claessen and van Wijk [15] analyzed axis-based methods and proposed an approach that allows the free configuration of these axes. This allows to select the most appropriate method for each attribute-attribute pair. We chose scatterplot matrices, because we need to investigate all combinations of attributes. Radar charts and parallel coordinate plots only display $n-1$ attribute combinations. The Hyperbox displays all combinations, but not in a uniform manner. Finally, the TimeWheel is especially suited for time-related data and puts a focus on the time axis, while we do not have such a distinguished axis.

Carr et al. [16] describe several scatterplot matrix techniques including binning, hexagonal binning (called hex-binning), density methods like symbols or contours, and perspective views. They discuss the use of sunflowers in using both square and hexagonal bins. We decided to use square bins, which in fact show interesting trends. The reason might be that sunflowers are more direction dependent and prone to reveal non-existing pattern than the tile approach presented here. Hao et al. [17] introduced variable binned scatter plots. They use non-uniform bins holding all values falling into that bin as single pixel. Values are ordered by position and category and all values are represented. They also allow for multiple levels of bins. Variable bins are not suitable in our case, as the general form and function of the scatterplot should be preserved. Moreover, a uniform bin size eases the comparison between and the linking among different scatterplots of the SPLOM. Finally, the data is too large, to be able to put all values as pixels on the screen, as this would use approximately one third of a two mega-pixel display.

Other approaches dealing with density are the approaches by Heinrich et al. [18], and Bachthaler and Weiskopf [19]. However, the focus of these methods is on the continuous rendering of the data. Our data is non-continuous.

Methods for navigating multi-dimensional spaces have been proposed by Asimov [20] and Elmqvist et al. [21] (scatterplot matrices).

## Methods

### Task

The visualization was developed based on the following task: given ≈ 800,000 data points with 11 dimensions, provide a visualization that allows to create and to verify hypotheses, as well as to gain additional information about the changes of the epigenetic marks during differentiation. For the concrete data set, 3 different methylations were used for creating a segmentation that is the basis for the analysis. Not only the comparison of the modifications among themselves but also the correlation with additional data such as length and CpG-density is of interest.

### Design Goals

The goals of our design are:

1  The visualization has to be flexible and should not assume specific properties of the data except a characterization of each data point with a certain number of dimensions. This requirement is inferred from the task described before.
2  Space efficiency
3  Good overview over the data
4  Recognition of pattern
5  Time efficiency (in generating the visualizations and in interaction)

### Tiled Binned Scatterplot Matrices

Scatterplots are well known visualizations for depicting the relationship between two attributes of a data set. Typically, the attributes have a continuous scale. A scatterplot consists of two axes: the x-axis and the y-axis. Each data point is mapped to a point where the x-position corresponds to the value of the first attribute and the y-position to the value of the second attribute, respectively.

A disadvantage of scatterplots is that overplotting might occur. Overplotting occurs, if data points have identical values (with respect to the scatterplot resolution) for the two attributes. This phenomenon occurs with a higher probability if the number of attributes or the number of data points is large. Binned scatterplots use so-called bins for collecting data points. Each axis is divided into intervals, the so-called bins. Each bin collects all the data points having values inside the border of the interval it represents. The amount of collected values can then be mapped to a visual variable, e.g., saturation, brightness, or transparency. The bins are represented by rectangles, which can be easily subdivided into smaller rectangles. Instead of positioning the data based on the attribute values, the bins are placed in the area between their lower and upper interval border. This fixes position and size at the same time. Bins represent the density of the data. Overall, binned scatterplots show the relationship between the attributes at bin resolution.

Each bin can be subdivided into sub-fields—called tiles—each representing a category of a categorical attribute. That is, each rectangle representing a bin is subdivided into smaller rectangles representing the tiles. Each category is assigned to a tile of the bin. Double encoding by position and another visual variable, e.g., color is possible.

If there are more than two attributes, scatterplot matrices (SPLOM) can be used to depict the correlations between all attribute pairs. To achieve this, each row and each column of the matrix represents one attribute. The cells of the matrix contain the scatterplot for the row-column attribute combination. The column attribute is mapped to the x-axis, while the row attribute is mapped to the y-axis.

As the diagonal would contain trivial scatterplots depicting the same attribute on the x- and the y-axis yielding only points on the main diagonal, histograms depicting the distribution for this attribute are shown in the cell of the SPLOM instead. These histograms are also binned and tiled. The x-axis represents the attribute value and is subdivided into bins. Each bin is subdivided again horizontally into tiles representing the categories of the categorical variable. The order of the tiles is the same as the order in the scatterplots. The y-Axis represents the number of data points falling into a bin for each category. We call a tiled binned scatterplot matrix TiBi-SPLOM.

### ChromatinScatter

*Visualization and Mapping*

In the chromatin scatter tool, a tiled binned scatterplot matrix (TiBi-SPLOM) is used for depicting the chromatin data. The data is split into two parts: the three ESC (embryonic stem cells) methylations are used to create a code in the range of zero to seven. The code is used as category that will be mapped onto the tiles. The methylations of MEF and NPC together with the CpG-density and the length of the segment form the 8 attributes that are shown in the scatterplot matrix (Figure 5).

The primary goal is to analyze the distribution of the original segments (ESC) based on their code in the other cell types, i.e., MEF and NPC. Using the tiled binned scatterplot matrices (TiBi-SPLOM), the ESC code is mapped onto the position and the color of the tiles. This double encoding helps finding pattern during analysis. A 3x3 subdivision of the bin is used yielding nine tiles, one more than ESC codes. The eight codes are assigned to the outer cells only, going from left to right and from top to bottom. Thus, the upper left cell represents code 0, the upper right cell represents code 2. The color is mapped using a colorblind safe and printer friendly color scheme from ColorBrewer [22] as shown in Table 1. The middle cell will always be a neutral white to minimize its influence on the other colors.

**Table 1** ESC Code (binary) to color mapping. Each row specifies the ESC code, the modifications in ESC, and the color, its name, and its RGB color.

| Code | Modifications | Color | Color Name | RGB color |
|------|---------------|-------|------------|-----------|
| 000 | none | | red | 227,26,280 |
| 001 | H3K9me3 | | bright green | 178,223,138 |
| 010 | H3K27me3 | | orange | 255,127,0 |
| 011 | H3K27me3, H3K9me3 | | bright blue | 166,206,227 |
| 100 | H3K4me3 | | blue | 31,120,180 |
| 101 | H3K4me3, H3K9me3 | | bright orange | 253, 191, 111 |
| 110 | H3K4me3, H3K27me3 | | green | 51, 160, 44 |
| 111 | H3K4me3, H3K27me3, H3K9me3 | | rose | 251, 154, 153 |

The number of bin elements is mapped to transparency. Two methods are used to scale the number of elements. The local scaling computes the relative amount of elements for one code in one bin divided by the total number of elements in this bin:

$$transparency(code, bin) = \frac{\#(code, bin)}{\#(bin)}$$

with *code* being the code of the chromatin segment, *bin* being the bin selected, $\#(code, bin)$ the number of elements with this code and in this bin, and $\#(bin)$ the number of elements in this bin. The global scaling computes the relative amount of elements for one code in one bin divided by the total number of elements having this code:

$$transparency(code, bin) = \frac{\log \#(code, bin)}{\log \#(code)}$$

with *code* being the code of the chromatin segment, *bin* being the bin selected, $\#(code, bin)$ the number of elements with this code and in this bin, and $\#(code)$ the number of elements having this code. For the global scaling, a logarithmic scale is used, as most elements fall into one bin.

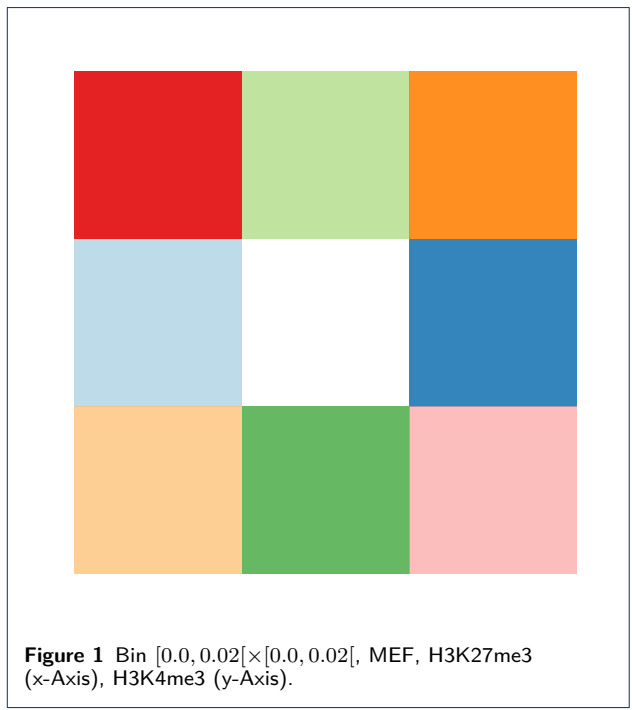

**Figure 1** Bin $[0.0, 0.02[\times[0.0, 0.02[$, MEF, H3K27me3 (x-Axis), H3K4me3 (y-Axis).

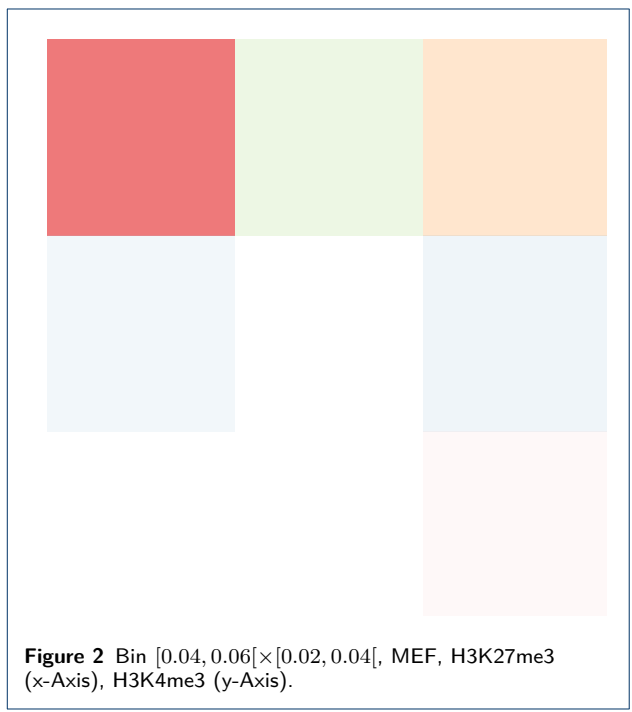

**Figure 2** Bin $[0.04, 0.06[\times[0.02, 0.04[$, MEF, H3K27me3 (x-Axis), H3K4me3 (y-Axis).

Two bins using the previously described position, color and transparency assignment are shown in Figure 1 and Figure 2. They are based on the scatter-plot shown in Figure 5 with 50 bins for each modification yielding a bin size of $0.02 \times 0.02$. For the bin $[0.0, 0.02[\times[0.0, 0.02[$ of the scatterplot showing the MEF modifications H3K27me3 (x-Axis) and H3K4me3 (y-Axis), all tiles are fully opaque, as most of the elements of MEF are unmodified, irrespectively of their original modifications in ESC (Figure 1). For the bin $[0.04, 0.06[\times[0.02, 0.04[$ of the same scatterplot, the tiles for the first four codes (0-3) are less opaque, codes 4 and 7 have a very high transparency, while codes 5 and 6 are completely transparent showing the background color white. The actual amount of modifications can be read from the information panel described below, showing that in fact no elements with codes 5 or 6 have been assigned to this bin.

Figure 3 shows the complete scatterplot for the MEF modifications H3K27me3 (x-Axis) and H3K4me3 (y-Axis). It is a zoomed version of the Scatterplot in Row 1, Column 2 of Figure 9. The annotations referred to in the subsequent discussion refer to the latter figure. Here, it can be seen that the diagonal, where the amount of modifications is equal for both modifications, is very opaque, i.e, a large amount of elements has been assigned to those bins (Figure 9 a). Essentially the same holds for the borders of the scatterplot, i.e., bins with no or complete modification for at least one of the modifications shown (Figure 9 b, c, d). A biological interpretation can be found in the "Results and Discussion" section.

Figure 4 shows the histogram for the MEF modification H3K4me3. The dark red lines are decreasing showing that most segments without modification stay unmodified. However, there is a relatively large amount of segments switching from unmodified to modified. On the other hand, the light red lines are increasing from left to right showing that most segments having this modification stay modified. These pattern give valuable insights to the biologists.

A complete example of a TiBi-SPLOM with global scaling is shown in Figure 5. The modification histograms show a so-called bathtub curve, the CpG-density histogram is left-skewed, and the length histogram shows that there are almost exclusively short elements with constant modification codes in ESC. The white space in the histograms with CpG and/or length show the same information. On the other hand, the histograms comparing two modifications show several visual effects:

1. The borders are comparatively dense, such that each histogram seems to have a border (Figure 9 a, b, c, d).

2. In the histograms between two modifications having the same cell type (MEF or NPC):

    (a) The main diagonal is visible (Figure 9 e).

    (b) Bins near the diagonal are less populated but still contain some elements.

    (c) Bins in the upper left (Figure 9 f) and the lower right corners tend to be sparsely populated or even empty—especially for the histogram comparing NPC H3K9me3 with H3K4me3.

The first two effects point to the conclusion that modifications prefer to be preserved not at all, completely, or having the same amount of modification.

The Biologists also showed interest in analyzing the data taking the length of the segments as reference. This led to adding a mapping of the length of the segments onto the position and the color of the tiles. As the length is originally a continuous variable, it was split into intervals (categories), which were mapped in turn onto the position and the color as shown in Table 2. For length, only five intervals (categories) were used. The transparency is calculated as described previously taking the length instead of the code as attribute. Both global and the local scaling can be used for length, too.

**Table 2** Length $l$ to color mapping. Each row specifies the length interval, the tile position (row, column) in the bin, and the color, its name, and its RGB color.

| Interval | Position | Color | Color Name | RGB color |
|---|---|---|---|---|
| $200 \le l \le 400$ | 1, 1 |  | red | 227,26,280 |
| $400 < l \le 600$ | 1, 2 |  | bright green | 178,223,138 |
| $600 < l \le 800$ | 1, 3 |  | orange | 255,127,0 |
| $800 < l \le 1000$ | 2, 1 |  | bright blue | 166,206,227 |
| $1000 < l$ | 2, 3 |  | blue | 31,120,180 |

### Interaction

The user can enter the height and the width of the TiBi-SPLOM. Further, the user can choose the number of bins for both x- and y-axis separately. Finally, it is possible to select the averaging method (local or global). and the displayed categorical attribute (ESC code or length).

Zooming (right mouse click) takes the histogram or scatterplot of the TiBi-SPLOM selected and displays it magnified as a single view. Selecting the same histogram or scatterplot again returns to the TiBi-SPLOM. Thereby, the initial position of the TiBi-SPLOM in the current viewport is preserved.

Selecting a bin (left mouse click) in the histogram or the scatterplot shows the following information (Figure 6):

- Attribute on the x-axis: number and descriptor
- Attribute on the y-axis: number and descriptor
- Bin on the x-axis: number, minimum value, and maximum value

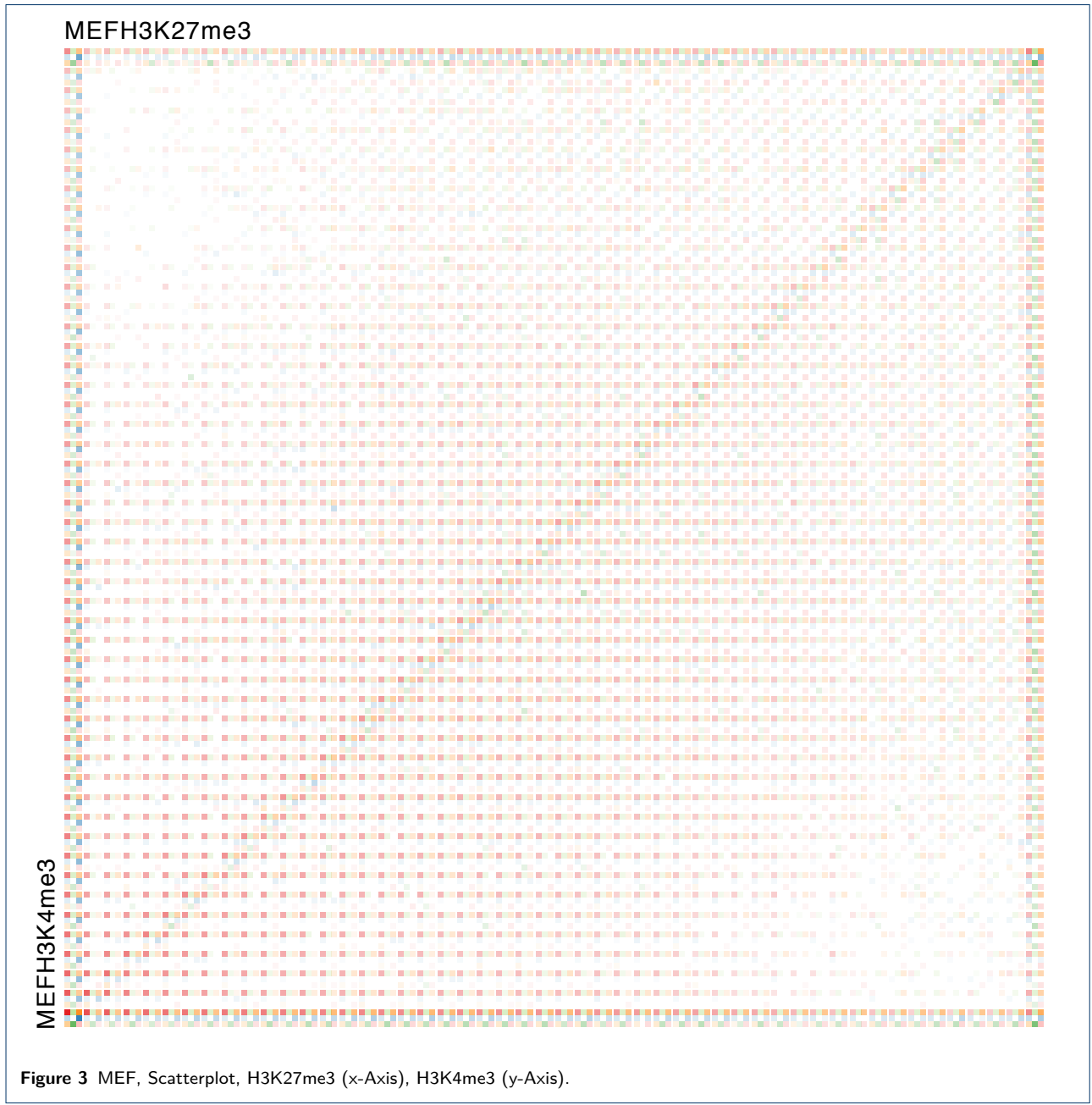

**Figure 3** MEF, Scatterplot, H3K27me3 (x-Axis), H3K4me3 (y-Axis).

- Bin on the y-axis: number, minimum value, and maximum value
- Averaging method: local or global average
- Color encoding: by ESC code or by length
- For each color (ESC code or length) are shown:
  - The color and the ESC code or the length interval, respectively
  - The color and the number of elements having this color and falling into the selected bin
  - The local maximum (number of elements in the selected bin)

- The global maximum (number of elements for this color—constant)

A range-slider filter [23] is used to select, which part of the data is shown in the scatterplot (Figure 7). The maximal range for these filters is

- $[0, 1]$ for modifications
- $[0, 1]$ for the CpG-density (normalized)
- $[0, \max(length)]$ for the length (normalized) using a logarithmic scale

Upon changing the filters, the minimum and maximum values of the axis in the TiBi-SPLOM are set to those chosen using the filters automatically, and the bins are recomputed using the new interval borders.

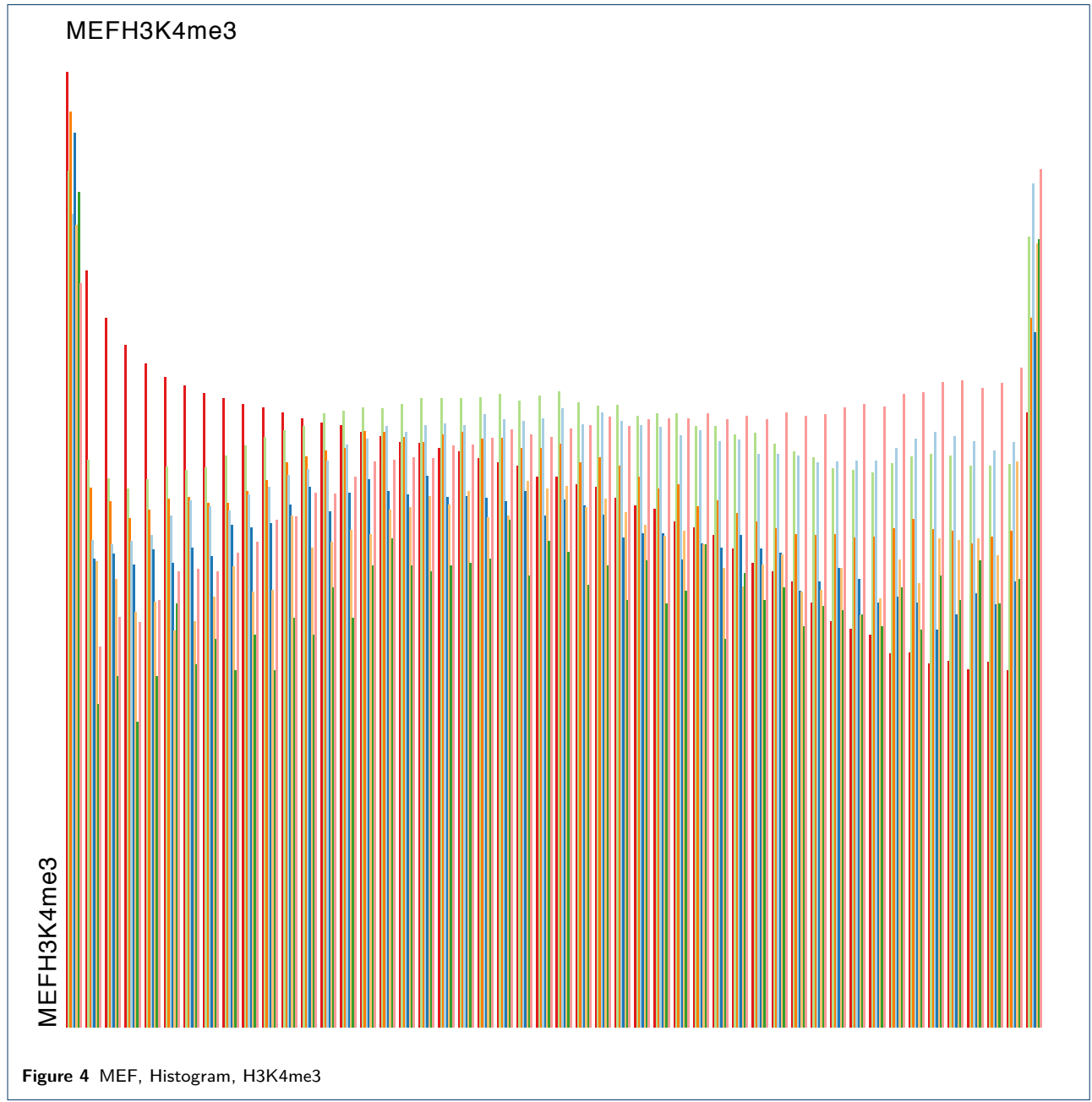

**Figure 4** MEF, Histogram, H3K4me3

The TiBi-SPLOM can be exported to different types of images such as png, jpg, bmp, gif, eps, pdf, or svg, or copied to the clipboard (only pixel-based file formats).

The chromatin scatter tool was implemented using Java 7.

### Design Properties

TiBi-SPLOMs are particularly well suited for this task, as they are flexible and not assuming specific properties of the data—fulfilling the design goal number 1.

Further, they provide a space efficient visualization of chromatin data (design goal number 2). White space in the visualization shows that the respective areas are empty and thus gives valuable information about the data distribution. They provide a good overview over the data (design goal number 3) as they summarize and abstract the data allowing high level analysis of the data's properties. Moreover, potentially interesting pattern are observable (see running example before and detailed discussion in the next chapter, design goal

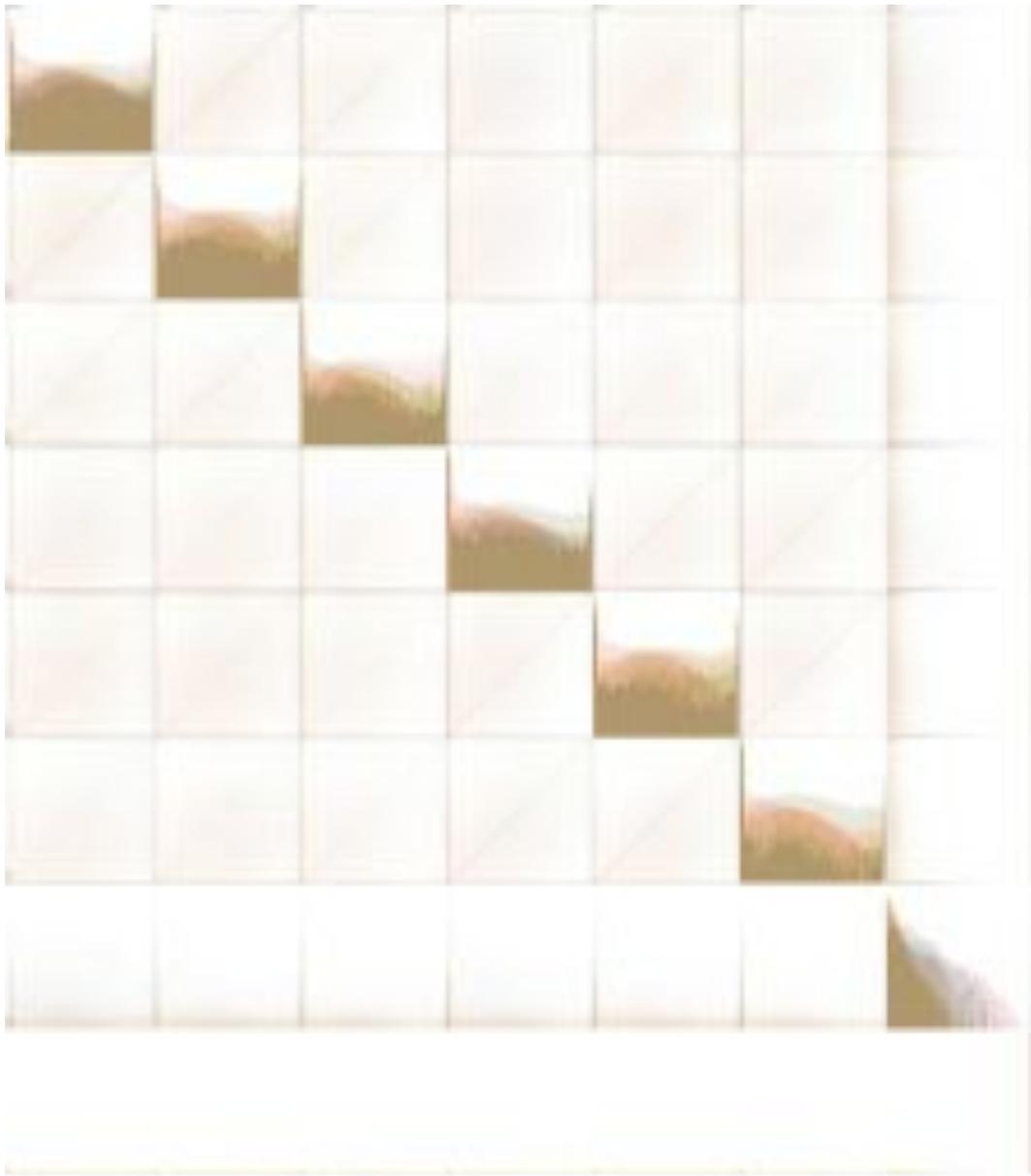

**Figure 5** TiBi-SPLOM showing MEF (H3K4me3 in row/column 1, H3K27me3 in row/column 2, H3K9me3 in row/column 3), NPC (H3K4me3 in row/column 4, H3K27me3 in row/column 5, H3K9me3 in row/column 6), CpG-density in row/column 7, and length in row/column 8 using global scaling based on the ESC codes. The modification histograms show a so-called bathtub curve, the CpG-density histogram is left-skewed, and the length histogram shows that there are almost exclusively short elements with constant modification codes in ESC. The white space in the histograms with CpG and/or length show the same information.

number 4). Finally, the visualizations are quickly generated and the interaction is fluent.

## Design Alternatives

The design alternatives for scatterplot matrices—other visualization techniques for multi-dimensional data—are presented and discussed in the previous work section. Here, the focus is on design alternatives regarding binning, tiling, and color mapping. Rectangular or square bins and tiles are a straight forward subdivision of scatterplots. Alternatives—like hexagonal bins—are not easily subdivided again. In particular, they can not be subdivided using hexagons only. A possible solution would be to use triangles for subdividing hexagons. However, the triangles subdividing a hexagon would have different orientations leading to different visual perceptions of data having identical relevance. Using circles for the bins would lead to additional white space reducing the visibility of the information. The same holds for using circles for the tiles.

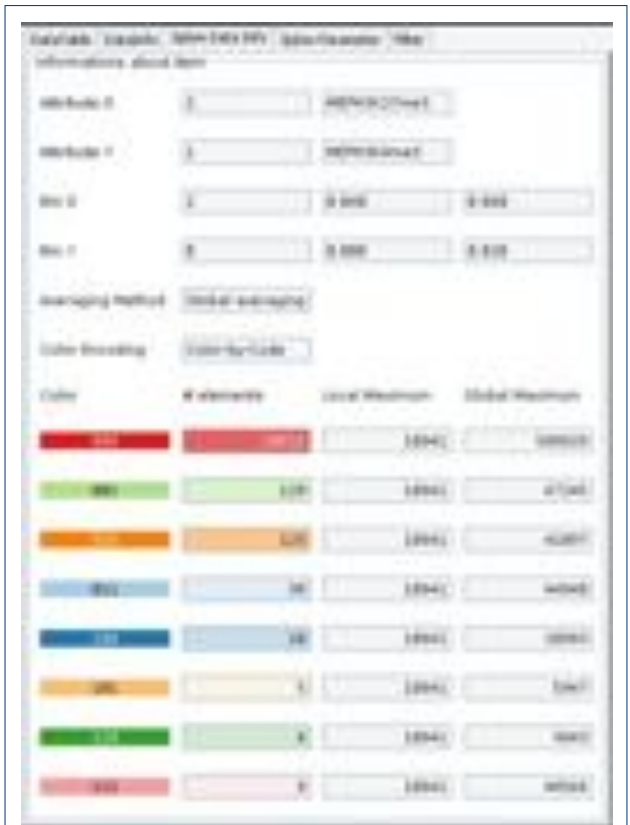

**Figure 6** Information shown upon selection of a bin from the TiBi-SPLOM.

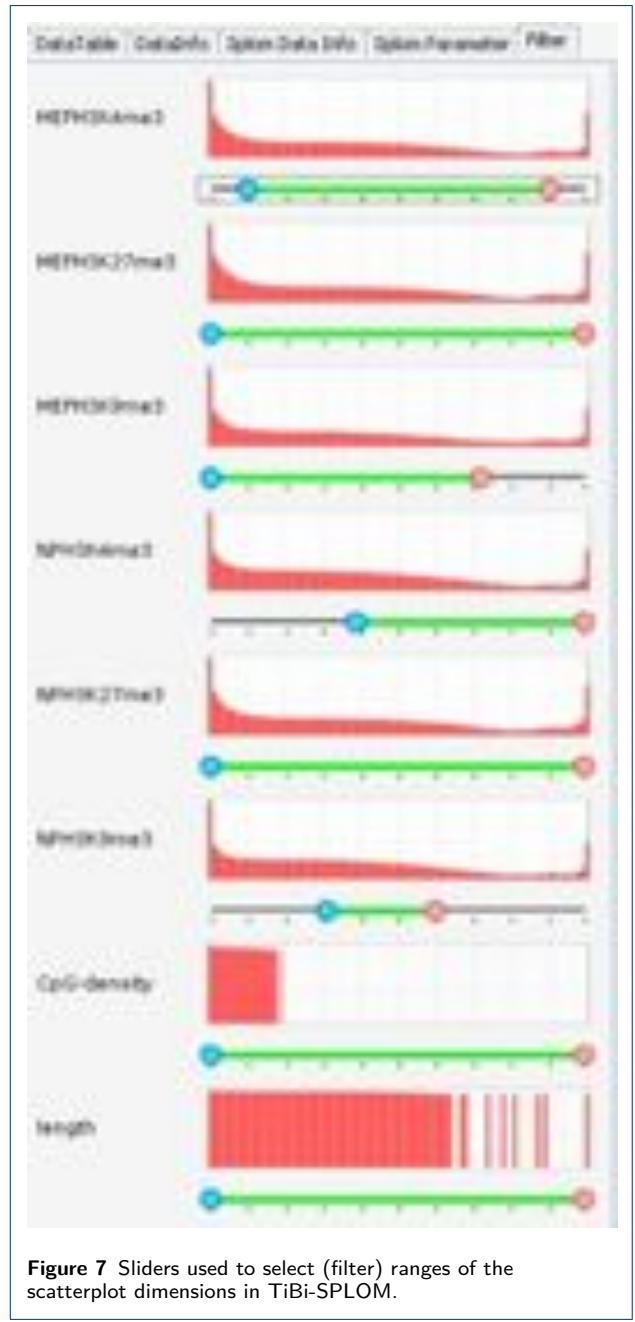

**Figure 7** Sliders used to select (filter) ranges of the scatterplot dimensions in TiBi-SPLOM.

The three single methylation appearances could be color encoded using red (ESC code 001), yellow (010), and blue (100). Then, the presence of exactly two methylations could be color encoded using the blending of the single methylation colors: orange (ESC code 011), purple (101), and green (110). However, a natural encoding for no methylation (ESC code 000) would be white and for three simultaneous methylations would be black (ESC code 111). Using saturation or transparency for density would lead to a gray scale for black, while it can not be used for white. Using other colors for ESC codes 000 and 111 is not possible, as for example brown, although being perceived as separate color, is just a dark shade of yellow. Moreover, no color choice for these codes would be intuitive and most colors would interfere with the six selected colors and transparency.

Transparency is used for encoding the bin size. Saturation would also be possible, however, it is already included in our color choice.

Alpha blending alone was not used as it necessitates scaling. Further, alpha blending for logarithmic scaling not easily achieved. Finally, alpha blending is computationally much more expensive than the method proposed here.

Texture or glyphs would require more space. Further, both can lead to visual artifacts that need to be overcome by, e.g., a hexagon grid [16]. However, the present encoding can be considered to be texture like as the tiles of the bin form the "bin texture". Using this en-coding, the pattern visible reveal correlations between the two dimensions of the individual scatterplots.

For the positioning—given the square subdivision of bins into tiles—only the assignment of the ESC code onto the position can be chosen freely. However, the influence of different positions on the analyzability will be minimal.

## Results and Discussion

The TiBi-SPLOMs were applied to the ES-segmentation data. The results for the global average with logarithmic scale are shown in Figure 5. Already, a coarse grained analysis of the TiBi-SPLOM reveals interesting pattern. In particular, the scatterplots and histograms in the 7th and 8th rows and columns look odd compared to the remaining plots.

At the first glance, the length histogram in the 8th row and 8th column of the scatterplot matrix in Figure 5 seems to have only four bars. Furthermore, most of the segments belong to the first bar. Only a few segments belong to the other bars, which seem to be composed of only segments with no modification in ESCs (red). However, the histograms, and also the scatterplots, depict the true range of the values. The maximal length is high but occurs only a few times, such that it cannot be visually detected in the histograms. The biological meaning of this observation is, that the data set includes some exceptionally long segments that are probably unmodified in ESCs. This also explains the appearance of the scatterplots between length and any other dimension—most of the segments are located in the first bin of the length dimension. Therefore, the white space in those plots highlights, that several combinations can not be found—e.g., exceptional long segments and high modification rate—by one of the modifications in the MEFs or NPCs. Since not much can be inferred from the current view, filtering is used to remove all segments with large length from the histograms and scatterplot allowing to obtain more interesting results presented below.

The 7th diagonal element of the TiBi-Splom in Figure 5 (shown enlarged in Figure 8) shows the distribution of the CpG-density. A large amount of data points have small CpG-density and any of the colors is abundant. The upper bound of CpG-density for which this is true is ≤ 0.015 (border marked with 1 in Figure 8), which can be easily obtained by picking the last bin with this property and getting the value from the Splom Data Info Panel (Figure 6). A CpG-density between 0.015 and 0.77 (between the borders marked with 1 and 2 in Figure 8) describes the middle level. This CpG-density level is usually modified with H3K4me3 (blue) or all three marks (rose) in ESCs. Less often but still frequent bivalently marked (green) or unmodified segments are found. The segments with high CpG-density (≥ 0.77, right side of border 2 in Figure 8) are not modified at all in ESCs (red), are marked with H3K4me3 (blue), or bivalently (green). This specific distribution of the different modifications states in ESCs to the CpG-density show that there are different recruitment mechanisms for modifiers of the three marks in ESCs.

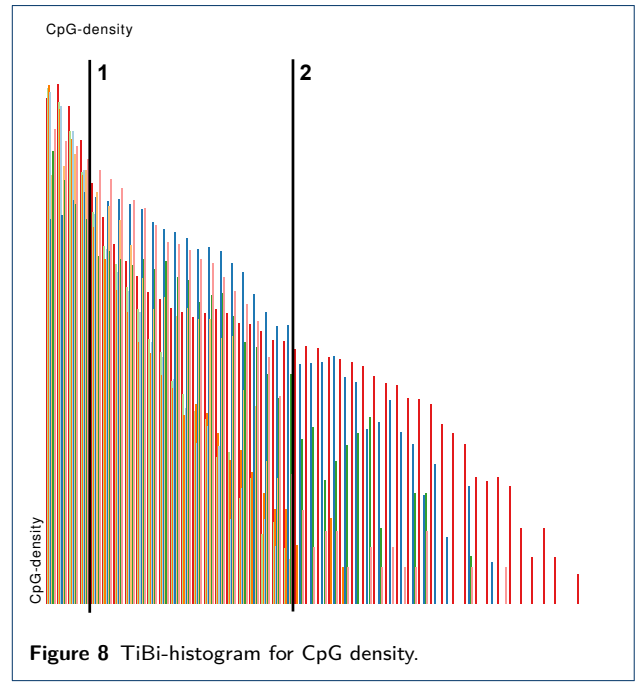

**Figure 8** TiBi-histogram for CpG density.

Similar to the scatterplots combining length with other dimensions, also the scatterplots including CpG-density and another dimension contain many uncolored and slightly colored tiles. This indicates that certain combinations of CpG-density and modification rate are rare or even not possible. We analyze the scatterplot for CpG-density and MEF H3K4me3 in detail below (Figure 10).

Histograms of the segment coverage by the different modifications in the different cell types yield strongly peaked distributions with peaks at 0 and/or 1. Thus, segments are either covered completely or not at all. Only a minority of the segments is covered partially with the modifications. Such a distribution would not be expected if modifications are distributed randomly within the genome. Even using a random distribution of modified segments in the genome, i.e., keeping the length distribution of the modified segments in the different cell types but randomly distributing them in the genome, does not lead to the observed distribution. Therefore, the assumption suggests itself that there is a mechanism defining those segments.

It was discovered that several modifiers act in a cooperative way, namely setting modifications preferen-tially in regions, which already carry this modification [24]. A propagation of the marks may be stopped by insulator proteins, which may serve as separator for the segments [25]. Furthermore, models show that modified regions are bistable and switch fast from completely modified to completely unmodified [26]. Bistability refers to the fact that segments stay modified even as the modification rates decrease when their segments are modified already. It has to decrease under a certain threshold to lead to demodification of the segments. On the other hand, exceeding this threshold of the modification rate does not lead to a modified segment if the segment was unmodified before. A much higher modification rate is required for modifying the segment in this case. Such a model is endorsed by our observation. Additionally, our observations give rise to the hypothesis that regions are modified in ESCs and stay modified until the cells lose the ability to keep them modified, e.g., due to changes in the ratio of modifying and demodifying processes (see [26]). Also, newly set modifications seem to stick to the regions defined in ESCs strongly.

In the following, some of the cells will be examined and analyzed.

### H3K9me3 in MEFs

The histograms on the diagonal of the TiBi-SPLOM provide information about a single modification, each. They allow analyzing state dependent changes regarding a reference. In our case, the reference is built out of the modifications in the embryonic stem cells (ESC), i.e., each color represents a certain combination of modifications in ESCs.

Looking at the histogram of H3K9me3 in MEFs (Figure 9, histogram in the bottom row, right column), one can recognize, as described above, the peaks at one and zero. Noticeably, most bars for the code representing the H3K9me3 modification in ESCs belong to the highest bars at the peak at zero and are even higher than the bar for the same code at the peak at one (compare bright blue, bright orange, bright green in peaks marked with 0 and 1). This indicates a massive loss of this modification when differentiating into MEFs. Thus, a lot of segments that were under repression in ESCs lost the repressive mark. As a consequence, genes on those segments may get under the control of marks, which switch on transcription. Furthermore, this is a sign for ESC specific silencing by H3K9me3.

Further analysis of the histogram and a detailed look at the distribution of the codes at the peak at one showed a dependency of H3K9me3 on the other marks. Most of the bars at the peak at one correspond to segments, which carry not only H3K9me3 but also one of the other marks while the bar with only H3K9me3 is relatively low (compare the bright green bar with the rose, bright blue, and bright orange in the region marked with 1). In particular, if segments were marked with all three modifications in ESCs (rose bar in peak 1), it stays modified during differentiation into MEFs. Thus, one can hypothesize, that the H3K9me3 modification is more stable if the same segment carries also other marks. There might be a cooperative effect on the different marks keeping segments with more than one mark modified throughout cell division. Whether this effect is mechanistic or represents the effect that demodifying those segments results in unfavorable or harmful states for the cell can not be determined yet.

The blue bar (H3K4me3 marked in ESCs) in peak 0 in the histogram for H3K9me3 in Figure 9 is as high as the red bar (unmodified in ESCs) in the same peak. In the other peaks of the histogram, the blue bar is very low. This shows that it is exceptional rare and hard to remove the activating mark H3K4me3 and replace it with the H3K9me3 mark when differentiating from ESCs into MEFs.

### Co-occurrence of H3K4me3 and H3K27me3 in MEFs

In Figure 9, scatterplot in the top row, middle column, the co-occurrence of H3K4me3 and H3K27me3 is shown. It exhibits a strong pattern is the distribution of the segments. The bins with highest number of segments can be found in the (0,0) and (1,1) corner of the scatterplot (mark a and b, respectively). This shows how strong both marks are correlated and supports the hypothesis proposed at the beginning. Either both marks are found completely on the segments or none of them.

Also, any border of the scatterplot as well as the diagonal are strongly populated with segments (regions a-e). The remaining parts of the scatterplot do not contain many segments and some of them are even empty, e.g. region f. The borders represent segments which carry only one of the two modifications. The diagonal means that there are several segments which contain both marks to the same amount. Interestingly, both, borders and the diagonal, are segments where the distribution of the segments in ESCs do not fit the distribution in MEFs, e.g. due to re-organization of the genome. These effects are not restricted to H3K4me3 and H3K27me3 but can be seen in any scatterplot comparing the modifications in one cell type. In comparisons of modifications between different cell types, this pattern cannot be observed which is easy to see in the full image of the TiBi-Splom in Figure 5. It is still unclear whether such a strong correlation of the marks, as can be seen in the diagonal, exists indeed in nature or whether this might be an artifact of the data.

Further analysis of the distribution of the codes revealed that segments which were modified with all three marks or two of three marks are more often found to be highly covered with both modifications than segments with just one or no mark in ESCs. Best see when comparing the intensity of the colors in regions a and b in Figure 9 where red, bright blue, orange, and blue are much more intense in a than in b while the other four colors are comparable in their intensity. Hypothesizing, that the stability of the marks is enhanced by a cooperative effect of not only the same modification but also other modifications. Thus, segments with two or three modification in ESCs are more likely to stay modified during differentiation into MEFs.

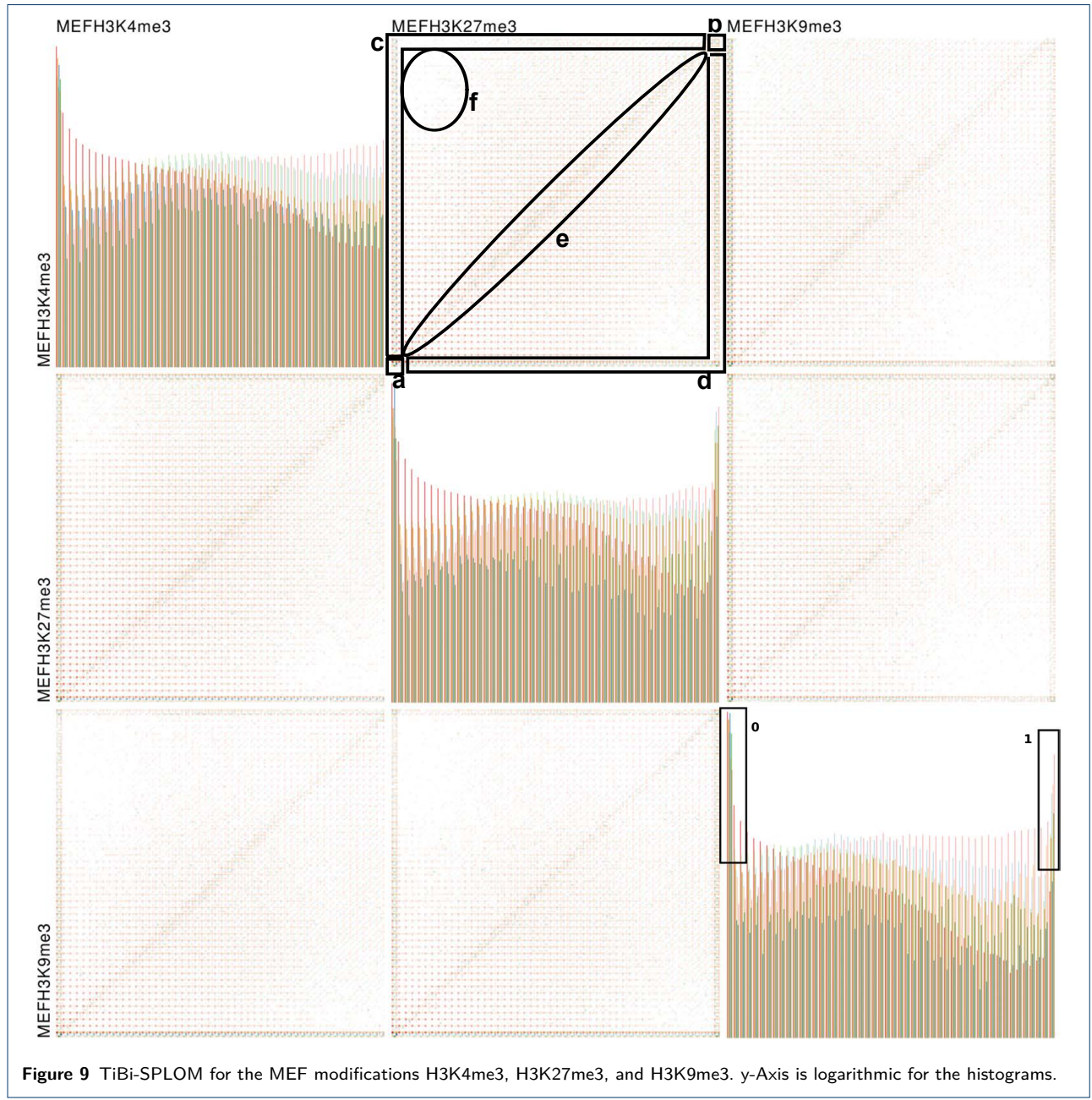

**Figure 9** TiBi-SPLOM for the MEF modifications H3K4me3, H3K27me3, and H3K9me3. y-Axis is logarithmic for the histograms.

### CpG-Density and H3K4me3 in MEFs

Additional to the modification data, the data set also contains CpG-density data. Literature often reports that there is a dependency of the modifying complexes on DNA motifs. In particular, H3K4me3 is found at the promoter for housekeeping genes which are often CpG-rich. Therefore, it is worthwhile to analyze the dependency of the modifications on the CpG-density. Figure 10 shows the scatterplot for the CpG-density (x-axis) and the H3K4me3 modifications in MEFs (y-axis). In the beginning of this section, the histogram was already analyzed an explained. It revealed that H3K4me3 is frequently found at low but also at mid-level CpG-density (blue and rose high at mid-level, any color is high at low CpG-density). However, at high CpG-density unmodified segments are much more frequent. Thus, H3K4me3 is dependent on CpG but favors certain CpG-density. This might be an effect of a specific CpG-containing motif.

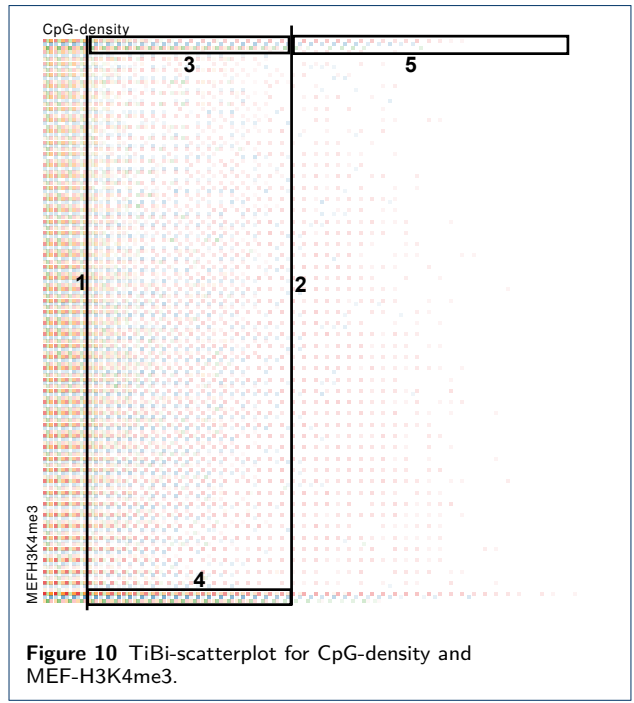

**Figure 10** TiBi-scatterplot for CpG-density and MEF-H3K4me3.

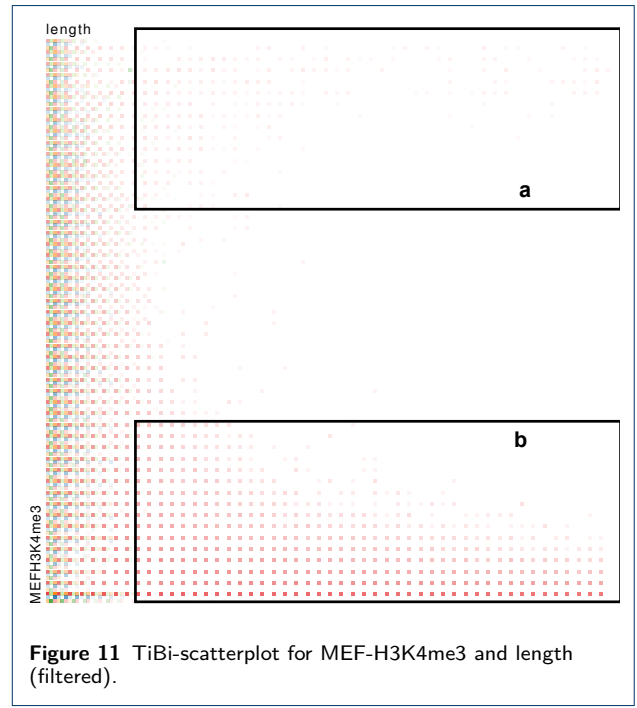

**Figure 11** TiBi-scatterplot for MEF-H3K4me3 and length (filtered).

The data proposed that there are 3 different levels of CpG-densities as seen in the histogram of the CpG-density (Figure 8). It is of interest whether the dependency on the recruitment mechanism changes. Therfore, we analyzed the scatterplot of CpG-density and H3k4me3 in MEF as an example (see Figure 10). The first level has CpG-densities lower than 0.015. It is enriched at any modification strength. The segments in this level are basically those that are set independently of the CpG-density. At a CpG-density between 0.015 and 0.77 (between border 1 and border 2), segments strongly tend to be either covered (region 3) or not covered (region 4) with H3K4me3, i.e., are localized at the top and the bottom of the scatterplot. Only a few are found between top and bottom (area between region 3 and 4). Many of them are depicted in blue (only H3K4me3 in ESCs) showing the dependency of H3K4me3 in ESCs. It seems that the CpG dependency of the modifying process gets lost in MEFs since the bottom line (region 4) of bins (no H3K4me3 in MEFs) contains more elements than the top line (region 3) independently of the color and thus, independently of the modification state in ESCs. Therefore, it is likely that the recruitment of the modifiers changes during differentiation into MEFs. At the highest CpG-density level ($> 0.77$), many red (code 000) segments are found having high coverage values for H3K4me3 in MEFs representing newly modified regions in MEFs (region 5).

From the data, the following hypotheses can be derived: The recruitment of H3K4me3 is not only dependent on high CpG-densities but is also strongly dependent on specific CpG-densities. Furthermore, the specific CpG-densities vary between the cell types. Thus, we may observe a switch in the recruitment of H3K4me3. This might be realized by a different composition of the complex trimethylation of the H3K4 in the nucleosomes binding to different motifs on the DNA.

### Length and H3K4me3

The length distribution showed that there are some very long segments. These segments are outliers of the length distribution and are unmodified in any cell type studied. Due to their existence, it is impossible to obtained insights into the length distribution of the modified segments. Therefore, we filtered by length such that even in the rightmost bin modified segments can be found.

We analyzed the correlation of H3K4me3 with the length in the cell types. The scatterplot for length and H3K4me3 in MEF is shown in Figure 11) and the one for length and H3K4me3 in NPC is shown in Figure 12). In MEFs, we marked two regions—a and b (Figure 11. While region 'a' consists of only rose color (all modifications set in ESCs), region 'b' consists of red color (unmodified in ESCs) solely. Furthermore, 'a' is located in the top part of the plot and 'b' in the bottom part. This means region 'a' contains segments, which are modified with H3K4me3 in MEFs and segments without the H3K4me3 marks in MEFs can be found in region 'b'. Both regions contain only long segments. Biologically, this means that long segments are either unmodified or modified with all three marks in ESCs (since no other color than rose is found for these lengths). If they carry all three marks in ESCs and are long, they will stay modified with at least H3K4me3 in MEFs. Long segments in ESCs, which are unmodified are not modified with H3K4me3 in MEFs. Thus, stability of the H3K4me3 mark may correlate with the length of the segments. This might be a combination of enzymes able to read and write H3K4me3 at the same time and in this way retains the long segments since it can benefit from the marks set on such segments.

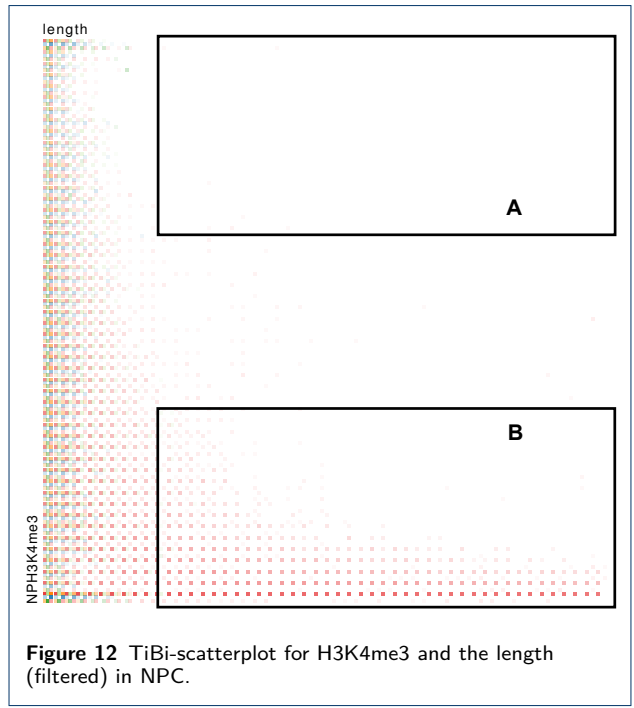

**Figure 12** TiBi-scatterplot for H3K4me3 and the length (filtered) in NPC.

We marked the region that corresponds to region 'a' and 'b' in MEF with 'A' and 'B' in NPC, respectively (see Figure 12). Comparing the image of MEF with the image of NPC, one can see that there is a difference: while there are rose points in region 'a', in region 'A' (same combination of length and H3K4me3 coverage but in NPC not in MEFs) there are no segments found. All long segments are now located in 'B'. Thus, the mechanism, which ensures that the long segments stay modified in MEFs, does not exist in NPCs. Consequently, the recruitment of the modifiers for H3K4me3 is fundamentally different in MEFs and NPCs.

# Conclusion

Tiled binned scatterplot matrices are efficient ways to represent multi-dimensional data and forster the analysis of the correlation between different dimensions. No prior knowledge of possible correlations in the data is required and they are highly usable to examine unknown data and formulate hypotheses.

Using them for analyzing an epigenetic data set with 11 dimensions (9 modifications, CpG-density, and length) led to two hypotheses. The first one regards the co-occurrence of marks and the formation of segments, which may be defined by so-called insulator proteins. The second hypothesis suggests different mechanisms to recruit modifying enzymes to their target position in the chromatin.

Future work comprises improving the interaction mechanisms and to continue the analysis of chromatin.

**List of abbreviations** TiBi-SPLOM: *tiled binned scatterplot matrix*, DNA: *deoxyribonucleic acid*, ChIP-seq: *chromatin immunoprecipitation sequencing*, NPC: *neuronal progenitor cells*, me3: *trimethylation*, H3: *histone H3*, K4: *lysine (one letter code K) at position 4*, H3K4me3: *trimethylation at histone H3 at the lysine at position 4*, H3K27me3: *trimethylation at histone H3 at the lysine at position 27*, H3K9me3: *trimethylation at histone H3 at the lysine at position 9*, CpG: *cytosine-guanine dinucleotide*.

**Competing interests**
The authors declare that they have no competing interests.

**Author's contributions**
DZ designed and developed the visualization. DG contributed to the visualization development. LS and SJP analyzed the results. All authors contribute to the manuscript writing. All authors read and approved the final manuscript.

**Acknowledgements**
LS and SJP were supported by the John Templeton Foundation Grant "Origins and Evolution of Regulation in Biological Systems" - ID: 24332. The opinions expressed in this publication are those of the author(s) and do not necessarily reflect the views of the John Templeton Foundation. We acknowledge support from the German Research Foundation (DFG) and Leipzig University within the program of Open Access Publishing.

**Author details**
[1]BSV, Leipzig University, Augustusplatz 10, 04109 Leipzig, Germany.
[2]InfAI e. V., Hainstraße 11, 04109 Leipzig, Germany. [3]Computational EvoDevo, Department of Computer Science, Leipzig University, Härtelstraße 16-18, 04107 Leipzig, Germany. [4]Center for Complexity & Collective Computation, Wisconsin Institute for Discovery, 330 North Orchard Street, WI 53715 Madison, USA.

**References**
1. Steiner, L., Hopp, L., Wirth, H., Galle, J., Binder, H., Prohaska, S.J., Rohlf, T.: A Global Genome Segmentation Method for Exploration of Epigenetic Patterns. PLOS ONE **7**(10), 46811 (2012)
2. Cheung, P., Lau, P.: Epigenetic regulation by histone methylation and histone variants. Mol Endocrinol **19**(3), 563–73 (2005)

3. Luger, K., Dechassa, M.L., Tremethick, D.J.: New insights into nucleosome and chromatin structure: an ordered state or a disordered affair? Nat Rev Mol Cell Biol **13**(7), 436–47 (2012)

4. Schübeler, D., MacAlpine, D.M., Scalzo, D., Wirbelauer, C., Kooperberg, C., van Leeuwen, F., Gottschling, D.E., O'Neill, L.P., Turner, B.M., Delrow, J., Bell, S.P., Groudine, M.: The histone modification pattern of active genes revealed through genome-wide chromatin analysis of a higher eukaryote. Genes Dev **18**(11), 1263–71 (2004)

5. Cao, R., Wang, L., Wang, H., Xia, L., Erdjument-Bromage, H., Tempst, P., Jones, R.S., Zhang, Y.: Role of histone H3 lysine 27 methylation in Polycomb-group silencing. Science **298**(5595), 1039–43 (2002)

6. Mikkelsen, T.S., Ku, M., Jaffe, D.B., Issac, B., Lieberman, E., Giannoukos, G., Alvarez, P., Brockman, W., Kim, T.-K., Koche, R.P., Lee, W., Mendenhall, E., O'Donovan, A., Presser, A., Russ, C., Xie, X., Meissner, A., Wernig, M., Jaenisch, R., Nusbaum, C., Lander, E.S., Bernstein, B.E.: Genome-wide maps of chromatin state in pluripotent and lineage-committed cells. Nature **448**(7153), 553–560 (2007)

7. Lehnertz, B., Ueda, Y., Derijck, A.A., Braunschweig, U., Perez-Burgos, L., Kubicek, S., Chen, T., Li, E., Jenuwein, T., Peters, A.H.: Suv39h-mediated histone H3 lysine 9 methylation directs DNA methylation to major satellite repeats at pericentric heterochromatin. Curr Biol **13**(14), 1192–200 (2003)

8. Bilodeau, S., Kagey, M.H., Frampton, G.M., Rahl, P.B., Young, R.A.: SetDB1 contributes to repression of genes encoding developmental regulators and maintenance of ES cell state. Genes Dev **23**(21), 2484–9 (2009)

9. Chernoff, H.: The use of faces to represent points in k-dimensional space graphically. Journal of the American Statistical Association **68**(342), 361–368 (1973)

10. Pickett, R.M.: Visual Analyses of Texture in the Detection and Recognition of Objects. In: Lipkin, B.S., Rosenfeld, A. (eds.) Picture Processing and Psychopictorics. Academic Press, Inc., Orlando, FL, USA (1970)

11. Inselberg, A., Dimsdale, B.: Parallel coordinates: A tool for visualizing multi-dimensional geometry. In: IEEE Visualization, pp. 361–378 (1990)

12. Alpern, B., Carter, L.: The hyperbox. In: IEEE Visualization, pp. 133–139 (1991)

13. Tominski, C., Abello, J., Schumann, H.: Axes-based visualizations with radial layouts. In: SAC, pp. 1242–1247 (2004)

14. Peng, W., Ward, M.O., Rundensteiner, E.A.: Clutter reduction in multi-dimensional data visualization using dimension reordering. In: INFOVIS, pp. 89–96 (2004)

15. Claessen, J.H.T., van Wijk, J.J.: Flexible linked axes for multivariate data visualization. IEEE Trans. Vis. Comput. Graph. **17**(12), 2310–2316 (2011)

16. Carr, D.B., Littlefield, R.J., Nicholson, W.L., Littlefield, J.S.: Scatterplot Matrix Techniques for Large N. Journal of the American Statistical Association **82**(398), 424–436 (1987)

17. Hao, M.C., Dayal, U., Sharma, R.K., Keim, D.A., Janetzko, H.: Variable binned scatter plots. Information Visualization **9**(3), 194–203 (2010)

18. Heinrich, J., Bachthaler, S., Weiskopf, D.: Progressive splatting of continuous scatterplots and parallel coordinates. Comput. Graph. Forum **30**(3), 653–662 (2011)

19. Bachthaler, S., Weiskopf, D.: Efficient and adaptive rendering of 2-d continuous scatterplots. Comput. Graph. Forum **28**(3), 743–750 (2009)

20. Asimov, D.: The grand tour: A tool for viewing multidimensional data. SIAM J. Sci. Stat. Comput. **6**(1), 128–143 (1985)

21. Elmqvist, N., Dragicevic, P., Fekete, J.-D.: Rolling the dice: Multidimensional visual exploration using scatterplot matrix navigation. IEEE Trans. Vis. Comput. Graph. **14**(6), 1539–1148 (2008)

22. Brewer, C.A.: Color Brewer. http://colorbrewer2.org/?type=qualitative&scheme=Paired&n=8#, accessed April 5th, 2014 (2014)

23. Williamson, C., Shneiderman, B.: The Dynamic HomeFinder: Evaluating Dynamic Queries in a Real-Estate Information Exploration System. In: SIGIR, pp. 338–346 (1992)

24. Hansen, K.H., Bracken, A.P., Pasini, D., Dietrich, N., Gehani, S.S., Monrad, A., Rappsilber, J., Lerdrup, M., Helin, K.: A model for transmission of the H3K27me3 epigenetic mark. Nat Cell Biol **10**(11), 1291–300 (2008)

25. Labrador, M., Corces, V.G.: Setting the boundaries of chromatin domains and nuclear organization. Cell **11**(2), 151–154 (2002)

26. Binder, H., Steiner, L., Przybilla, J., Rohlf, T., Prohaska, S., Galle, J.: Transcriptional regulation by histone modifications: towards a theory of chromatin re-organization during stem cell differentiation. Physical Biology **10**(2), 026006 (2013)